\documentclass[%
 reprint,
 amsmath,amssymb,
 aps,
]{revtex4-1}

\usepackage{graphicx}
\usepackage{dcolumn}
\usepackage{bm}

\begin{document}

\preprint{APS/123-QED}

\title{Generic Adaptation by Fast Chaotic Exploration and Slow Feedback Fixation}

\author{Yuuki Matsushita}
\affiliation{
Department of Biological Sciences, Graduate School of Science, Osaka University, Machikaneyama-cho, Toyonaka, Japan
}
\author{Kunihiko Kaneko}
\affiliation{
Center for Complex Systems Biology, Universal Biology Institute, University of Tokyo, Komaba, Tokyo 153-8902
}
\affiliation{
The Niels Bohr Institute, University of Copenhagen, Blegdamsvej 17, Copenhagen, 2100-DK, Denmark
}


\date{\today}

\begin{abstract}
Living systems adapt to various environmental conditions by changing their internal states.
Inspired by gene expression and epigenetic modification dynamics, we herein propose a generic mechanism for adaptation by combining fast oscillatory dynamics and a slower feedback fixation process.
Through extensive model simulations, we reveal that fast chaotic dynamics serve as global searching for adapted states fixed by slower dynamics.
The mechanism improves as the number of elements is increased.
Relevance to cellular adaptation and optimization in artificial neural networks is also discussed herein.
\end{abstract}

\maketitle
Biological systems can generally adapt to various environmental conditions by adjusting their internal states, and this is essential to their survival and universal for living systems.
The most investigated and established mechanism for adaptation is signal transduction networks in cells, by which the information on external conditions alters the gene expression pattern to fit the environments.
Such networks, which are considered to be evolved over generations to meet given environmental conditions, have been explored in the literature \cite{Alberts2017-sz}.

Even though such a signal transduction mechanism is important and has been thoroughly investigated, it remains unclear as to whether it can explain all cell adaptations \cite{Braun2015-wy,Kashiwagi2006-px,Ciechonska2022-kl, Tsuru2011-mt}.
Braun, for instance, demonstrated that yeasts can adapt to various conditions, including de novo conditions that their ancestors have not experienced \cite{Braun2015-wy}.
Such spontaneous adaptation has been observed in artificially embedded networks without environmental information \cite{Kashiwagi2006-px}.
Uncovering some alternatives---possibly generic mechanisms, if any---is required.

As a possible mechanism, attractor selection, wherein a state with a higher growth rate is selected by taking advantage of noise and growth dilution, was proposed \cite{Furusawa2008-eo, Furusawa2013-pc}.
Despite the generality and applicability of the mechanism \cite{Leibnitz2010-nf,Koizumi2010-qw}, the existence of attractors in gene expression dynamics fitted to the environment need to be provided.
As a remedy for such demand, an introduction of the epigenetic modification process, which could generate different stable states from the original gene expression dynamics, was proposed \cite{Furusawa2013-pc}.
This can enhance the applicability of the mechanism; however, an evolutionary process to optimize the network is further needed \cite{Furusawa2013-pc, Gombar2014-rv}.

Inspired by these arguments, we herein propose an alternative and generic adaptation mechanism not restricted to cells but applicable to optimization problems in general.
We adopt oscillatory (chaotic) dynamics, stimulated by recent studies in cellular differentiation \cite{Matsushita2020-fy,Matsushita2022-rx},
in which epigenetic modification leading to robust cellular differentiation and reprogramming is proposed \cite{Matsushita2020-fy, Matsushita2022-rx}.
Here, epigenetic modifications are biomolecular mechanisms such as DNA methylation or histone modification \cite{Bird2007-tw, Cortini2016-vu}.
Even though their detailed mechanisms are different \cite{Cortini2016-vu,Hihara2012-ga,Tripathi2019-de}, they would generally change the feasibility of gene expression, which can alter the gene expression dynamics and their stable expression patterns.
In the proposed theory for differentiation, the interplay between oscillatory gene expression dynamics and slower epigenetic modification generates and stabilizes novel cell types.

In this letter, we examine whether such interplay between gene expression dynamics and slower positive feedback can generally work for adaptation.
Through the oscillatory dynamics, possible states are explored.
Then, if an adapted state is reached, the slow (epigenetic) process will work efficiently to fix such a state.
We provide a simple model of (gene) regulatory networks that can produce oscillatory dynamics with a positive (epigenetic) feedback process to demonstrate such a mechanism.
Via extensive simulations of the model, we show that the model can adapt to a variety of external conditions if the original dynamics show sufficiently complex (chaotic) oscillation.
We obtain the condition to achieve generic adaptation and demonstrate that the fraction of networks satisfying the condition increases with the number of units (genes).
We discuss the generality of the mechanism, including the application to machine learning and optimization in neural and artificial networks.

We consider a model consisting of \( N \) genes with a regulatory network and slower epigenetic modification.
The \( i \)-th unit (gene) has variables \( x_i \) and \( \theta_i \) ( \( i = 1, 2, \dots, N \)), where \( x_i \) represents the \( i \)-th (gene) expression level (concentration of the protein corresponding to the gene) and \( \theta_i \) represents the epigenetic modification level of the \( i \)-th gene 
\footnote{
Here, we describe the model in terms of genes; however, it can be generalized to introducing units with a slow feedback fixation process.
}.
Genes activate or suppress each other via synthesized proteins according to gene regulatory matrix \( J_{ij} \).
If \( J_{ij} \) is positive (negative), the \( j \)-th gene activates (suppresses) the \( i \)-th gene \cite{Mjolsness1991-vr,Salazar-Ciudad2000-ov,Salazar-Ciudad2001-qt} as given by
\begin{align}
        \frac{d x_i}{d t} &= F \left( \frac{1}{\sqrt{N}} \sum_j J_{ij} x_j + \theta_i \right) - x_i, \label{eq:dx}
\end{align}
where \(F(z) \) is a monotonic function exhibiting an on-off switch, as \( F(z) = \tanh (\beta z) \).
We set \( \beta = 40 \), i.e., \( F(z)\) is close to a step function.
The value of \( x_i = 1\) or \( -1 \) represents full- or non-expression of the \( i \)-th gene.
In Eq. \eqref{eq:dx}, \( -\theta_i \) works as a threshold for the expression of the \( i \)-th gene.
In other words, the epigenetic modification level \( \theta_i \) gives the feasibility of the \( i \)-th gene expression.
As \( \theta_i \) increases (decreases), less (more) input from other genes is required for expression.
For epigenetic modification dynamics, we adopt the simplest form of reinforcement \cite{Furusawa2013-pc, Miyamoto2015-ux, Matsushita2020-fy, Huang2020-yk, Matsushita2022-rx}:
\begin{align}
    \frac{d \theta_i}{d t} &= v^k(t) (x_i - \theta_i), \label{eq:dtheta}
\end{align}
where \( v^k(t) \) is positive and indicates positive feedback between gene expression and epigenetic modification.
If the \( i \)-th gene is expressed, it is more feasible to be expressed following the increase in \( \theta_i \).
Such positive feedback is based on previous experimental results \cite{Schreiber2002-gm, Dodd2007-kz, Sneppen2008-ui, Hihara2012-ga}.
Here, \( v^k(t) \) is the timescale of the epigenetic modification process depending on cellular fitness.
It is set always smaller than unity, that is, change in \( \theta_i \) is slower than that in \( x_i \).
Within this range, \( v^k(t) \) is increased when the cell is more fitted to the \( k \)-th environmental condition.

The fitness is given by the expression pattern of output genes \( x_m (m = 1, 2, \cdots, M < N)\).
By introducing \( X_m^k \) as a target (desired) gene expression pattern under the \( k \)-th environment, fitness is given by the distance between \(x_m \) and \( X_m^k \) according to
\begin{align}
    \mathrm{fitness}^k \equiv \sqrt{ \sum_m^M (x_m(t) - X_m^k)^2/M}. \label{eq:fitness}
\end{align}
Then, \( v^k(t) \) is determined as
\begin{align}
v^k(t) = v^{\mathrm{max}} \exp \left\{ -b\times(\mathrm{fitness}^k) \right\}, \label{eq:v}
\end{align}
where \( b = 4\) and \( v^{\mathrm{max}} = 10^{-1} \).
Note that \( v^k(t) \) takes the maximum value \( v^{\mathrm{max}}\) when \( x_m \) is equal to \( X_m^k \), in which case cell completely adapts to the \( k \)-th environment.

We herein adopt \( M = 5 \).
Then, the total number of possible \( M \)-bit target patterns with \( -1\) or \( 1 \), described as \( \{ 1, 1, 1, 1, 1 \}, \{-1, 1, 1, 1, 1, 1\}, \dots, \{ -1, -1, -1, -1, -1 \} \), is \( 2^M = 32 \).
However, the present cell model notably has a symmetry \( \boldsymbol{x} \leftrightarrow - \boldsymbol{x} \).
Considering this symmetry as \( \boldsymbol{X}^k \) and \( - \boldsymbol{X}^k \), there are \( 2^M /2 = 16 \) independent \( M \)-bit patterns \( \{ \boldsymbol{X}^k \} \).
Each environmental condition \( k = 1, 2, \cdots, 16 \) has the corresponding target expression pattern \( \boldsymbol{X^k} \); the adaptation to each of these \( M \)-bit patterns are examined.

We adopt random gene regulatory networks (GRNs), whose elements \( \{ J_{ij} \} \) are randomly assigned either \( \pm 1, 0 \), with equal probability.
Each of adaptation trials is started from \( \theta_i = 0 \) state and randomly chosen \( x_i \).
Each of the targets \( \boldsymbol{X}^k (k = 1, 2, \cdots, 16) \) is assigned for each trial, and we run the dynamics of Eqs. \eqref{eq:dx}-\eqref{eq:v} until they reach the final stationary state.
If the final \( x_m \) is equal (or sufficiently close) to \( X_m^k \), the cell adapts to the \( k \)-th environment.

\begin{figure}[tbp]
    \centering
    \includegraphics[width=\linewidth]{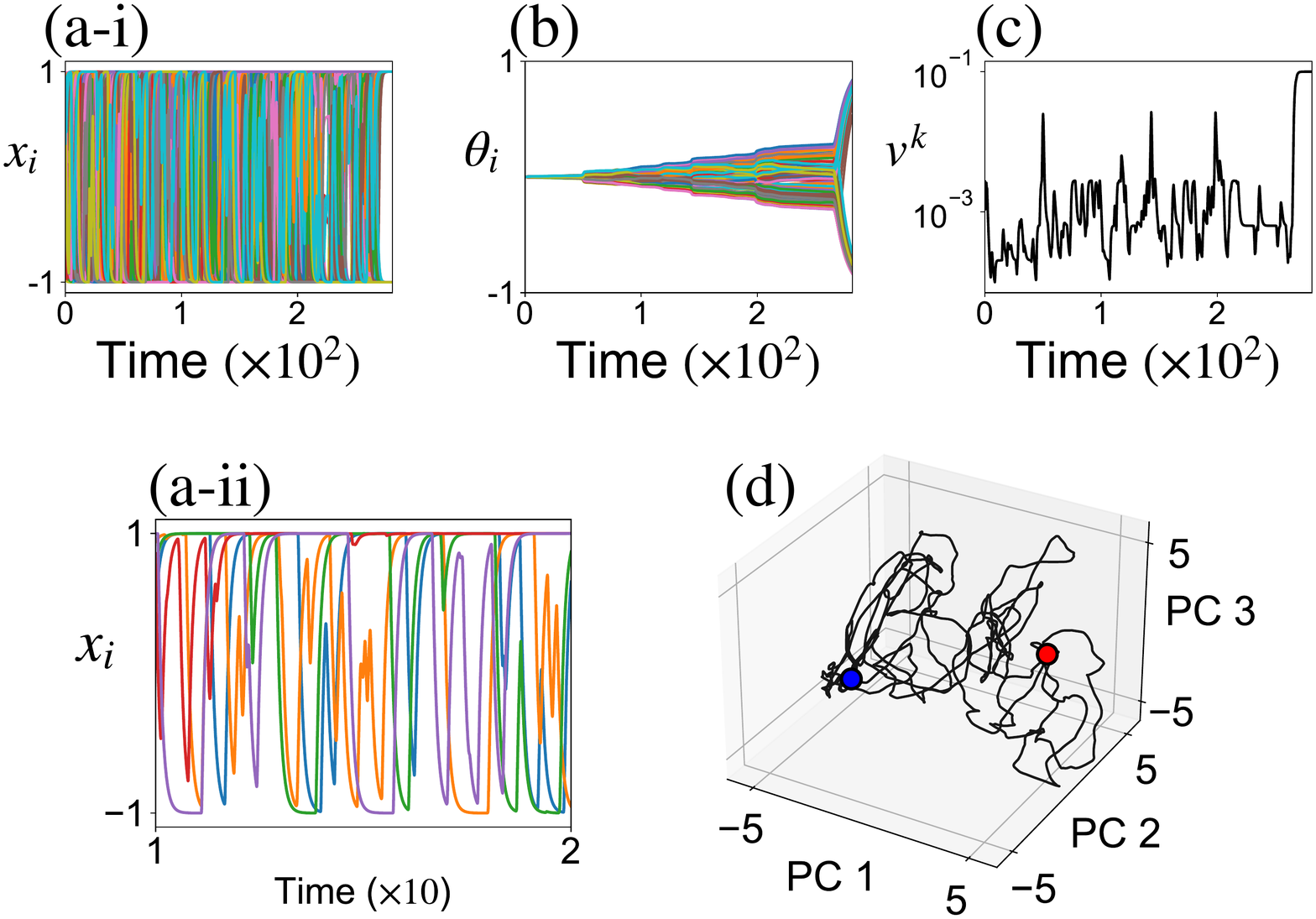}
    \caption{
        Adaptation of cell model with \( N = 100 \).
        Time series of \( x_i (i = 1, 2, \dots, N) \)(a-i, ii), \( \theta_i \)(b), and \( v^k \)(c).
        Starting from the initial condition with random \( x_i \) and \( \theta_i = 0 \), the cellular state converges to transient gene expression oscillation.
        With gradually developed \( \theta_i \), the cellular state reaches desired gene expression pattern \( X_m^k \).
        That is, \( v^k \) takes the maximum value \( v^{\mathrm{max}}\) (\( = 10^{-1} \)).
        (a-ii) Time-series of \( x_i \) only for \( i = 1, 2, \dots, 5\) for \( t = 10 \sim 20\).
        (d) Adaptation dynamics in \( \boldsymbol{x} \) space.
        We adopt PCA obtained from oscillatory dynamics with \( \theta_i = 0 \).
        The cellular state starting from a random initial condition (red X) reaches the target gene expression pattern (blue point) throughout transient oscillatory dynamics.
        }
        \label{Fig1}
\end{figure}

Figure \ref{Fig1} shows the time series of \( x_i \), \( \theta_i \), and \( v^k \) in success adaptation to a certain environment.
In Fig. \ref{Fig1}(a), gene expression dynamics \( x_i \) first converge to an irregularly oscillating state.
Then, the oscillatory dynamics are fixed by epigenetic modification \( \theta_i \) when \( x_m \) approaches \( X_m^k \) and the fitness \( v^k(t) \) is increased (Fig. \ref{Fig1}(b)).
Finally, throughout these transient dynamics, the cellular state adapts to the desired target state when \( x_m \) reaches \( X_m^k \).
The value of \( v(t) \) varies in time at first (Fig. \ref{Fig1}(c)), reflecting transient dynamics of \( x_i \), until a fitted state is selected.
Figure \ref{Fig1}(d) shows the adaptation dynamics in \( \boldsymbol{x} \) space using principal component analysis (PCA) obtained from oscillatory dynamics with \( \theta_i = 0 \).

\begin{figure}[tbp]
        \centering
        \includegraphics[width=0.75\linewidth]{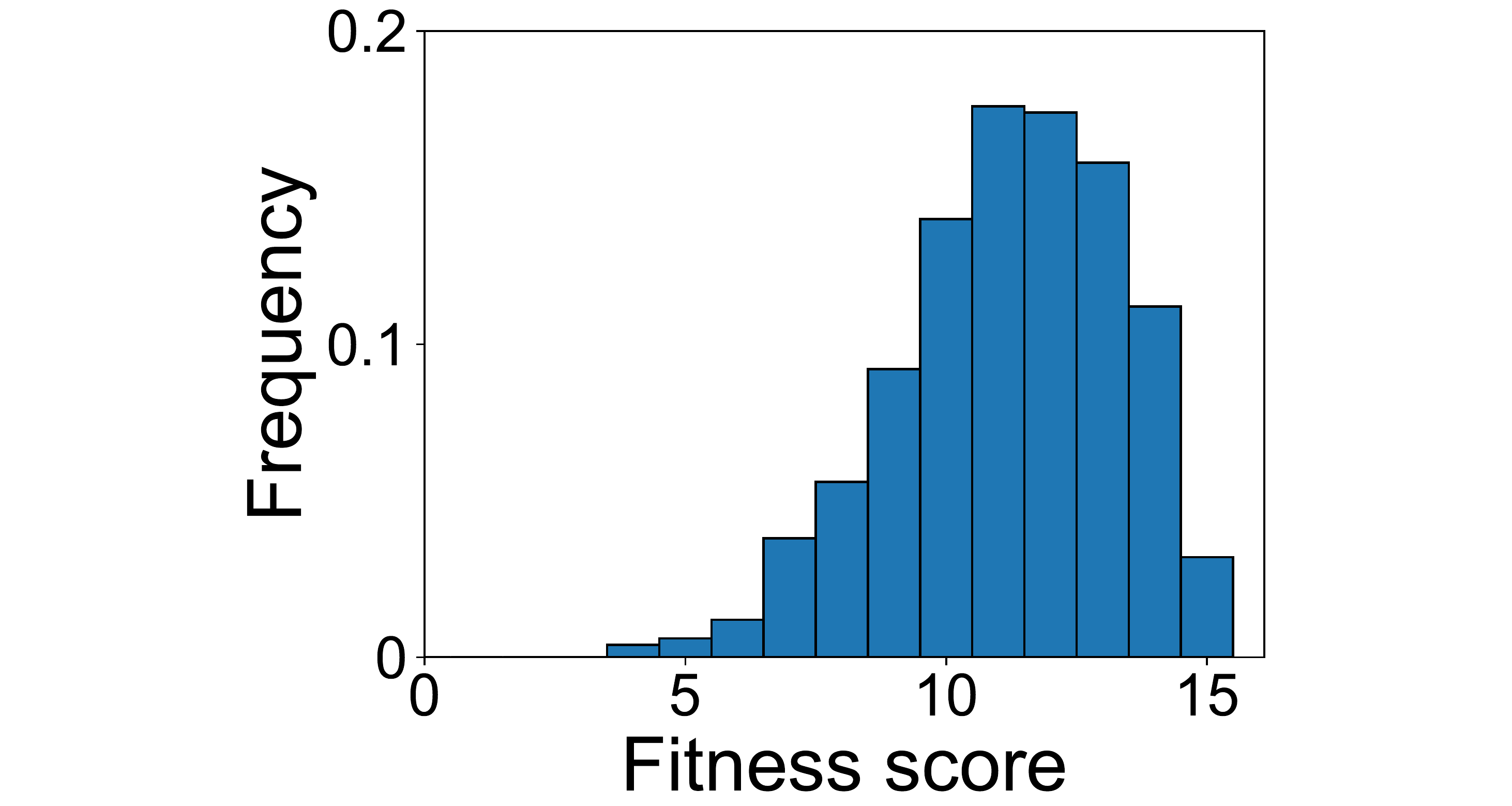}
        \caption{
        Distribution of adaptation capacity (= fitness score) computed from 500 random network models with \( N = 100 \).
    }
    \label{Fig2}
\end{figure}

Next, the adaptation capacity of the cell is investigated.
The number of environments to which adaptation is achieved among 15 environmental conditions gives the fitness score, i.e., adaptation capacity.
Figure \ref{Fig2} shows the distribution of the capacity over 500 random GRNs with \( N = 100 \).
Here, the criterion of adaptation to the \( k \)-th environment is given by if the trial in the \( k \)-th environment finishes with \( |x_m - X_m^k|_2 < 10^{-2} \) at least for one out of three trials.
For most of the random regulatory networks \( \{ J_{ij} \} \), cells can adapt to more than half of the \( 16 \) environments.

\begin{figure}[tbp]
    \centering
        \includegraphics[width=\linewidth]{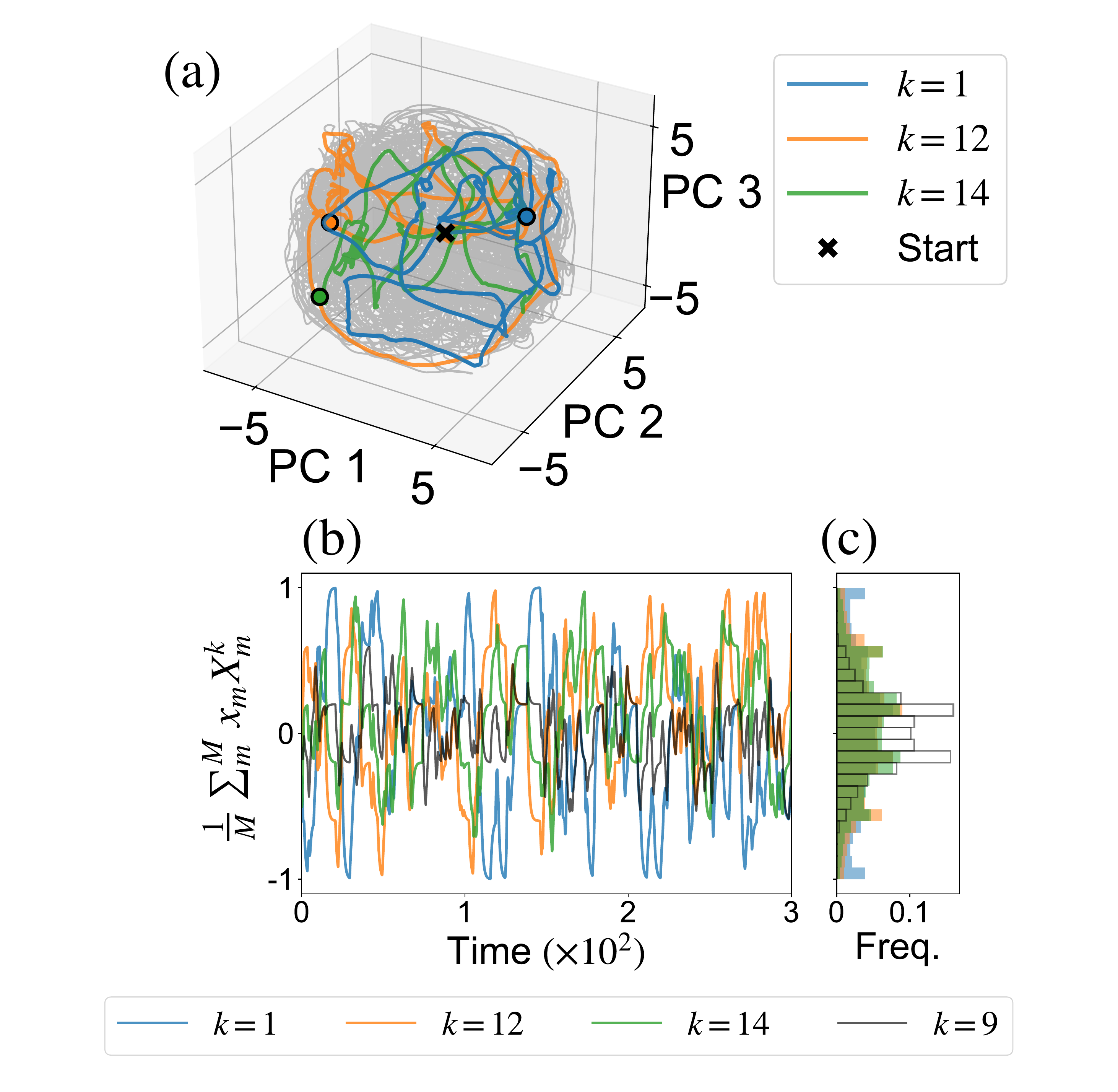}
    \caption{
        (a) Adaptation process against three different environmental conditions (targets) plotted in \( \boldsymbol{x} \) space using the PCA space for oscillatory dynamics with \( \theta_i = 0 \) (gray trajectory).
        The colored trajectories show adaptation in three different environments (\( k = 1, 12 , 14 \)) starting from identical initial condition (X).
        Each adaptation is completed at the colored circle.
        (b) Time series (left) and histogram (right) of the overlap of \( \{ x_m (t) \} \) with \( \theta_i = 0 \) and target patterns \( (1/M) \sum_m^M x_m X_m^k \).
        Black time series and histogram represent the case of non-adaptable environments (\( k = 9 \)).
    }
    \label{Fig3}
\end{figure}

Next, the adaption of cells to multiple environmental conditions is examined.
Figure \ref{Fig3}(a) shows the dynamics in \( \boldsymbol{x} \) against three different environmental conditions from identical initial condition by adopting PCA obtained from oscillatory dynamics with \( \theta_i = 0 \).
In Fig. \ref{Fig3}(a), the gray curve corresponds to the trajectory with \( \theta_i = 0 \), which shows chaotic dynamics.
With epigenetic modification dynamics of Eq. \eqref{eq:dtheta}, the fitted state is reached and fixed after transient (chaotic) oscillation, depending on each target condition \( k \).
In Fig. \ref{Fig3}(b), we study how \( \{ x_i(t) \} \) with \( \theta_i = 0 \) and each of target patterns \( \{ \boldsymbol{X^k} \} \) come closer by introducing the inner product of \( \boldsymbol{x} \) and \( \boldsymbol{X^k} \) given by \( (1/M) \sum_m^M x_m X^k_m \) to characterize the distance between \( \boldsymbol{x} \) and the \( k \)-th target pattern.
As shown in Fig. \ref{Fig3}(b), the time series with \( \theta_i = 0 \) explores globally the phase space and approaches the target patterns that can be adapted (\( k = 1, 12, 14 \) in this example); however, it cannot approach target that cannot be adapted (\( k = 9 \)), where the inner product remains around \( 0 \).
The distribution for inner products is extended globally over \( [-1, 1] \) for the former case but is centered around zero for the non-adapted case (Fig. \ref{Fig3}(c)).

\begin{figure}[tbp]
        \includegraphics[width=\linewidth]{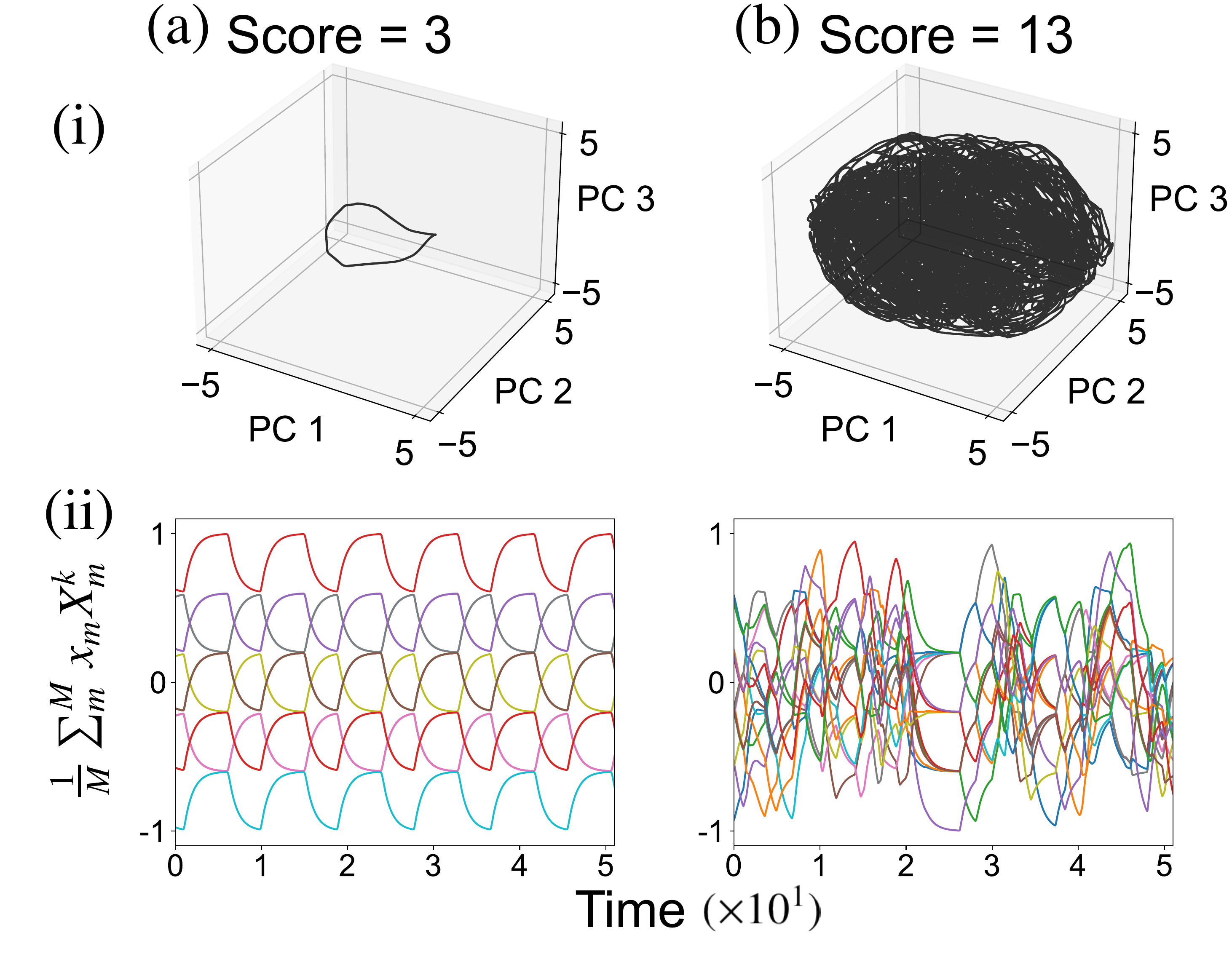}
    \caption{
        Comparison of the dynamics with fixed \( \theta_i = 0 \) for small (score = 3) and large adaptation capacity with \( N = 60 \).
        (a) Dynamics plotted in the PCA space of \( \boldsymbol{x} \).
        (b) Time series of \( (1/M) \sum_m^M x_m X_m^k \) and the overlap of \( \{ x_i(t) \} \) with \( \theta_i = 0 \) and target patterns \( \{ \boldsymbol{X}^k \} \).
        Left: Score = 3, globalness = 0.16, and no positive Lyapunov exponents.
        Right: Score = 13, globalness = 0.35, and two positive Lyapunov exponents 0.51, 0.16.
    }
    \label{Fig4}
\end{figure}

We next focus on how gene expression dynamics with \( \theta_i = 0 \) depend on the adaptation capacity.
Figure \ref{Fig4}(i) shows the dynamics for small (score = 3, (a)) and large (score = 13, (b)) adaptation capacities.
Comparing these trajectories, gene expression dynamics with small adaptation capacity travel small portions of phase space, whereas those with large capacity travel large portions of phase space (Fig. \ref{Fig4}(ii)).
In Fig. \ref{Fig4}(i), the former has a limit cycle attractor (a) while the latter has a chaotic attractor with two positive Lyapunov exponents (b).

\begin{figure}[tbp]
    \centering
    \includegraphics[width=0.9\linewidth]{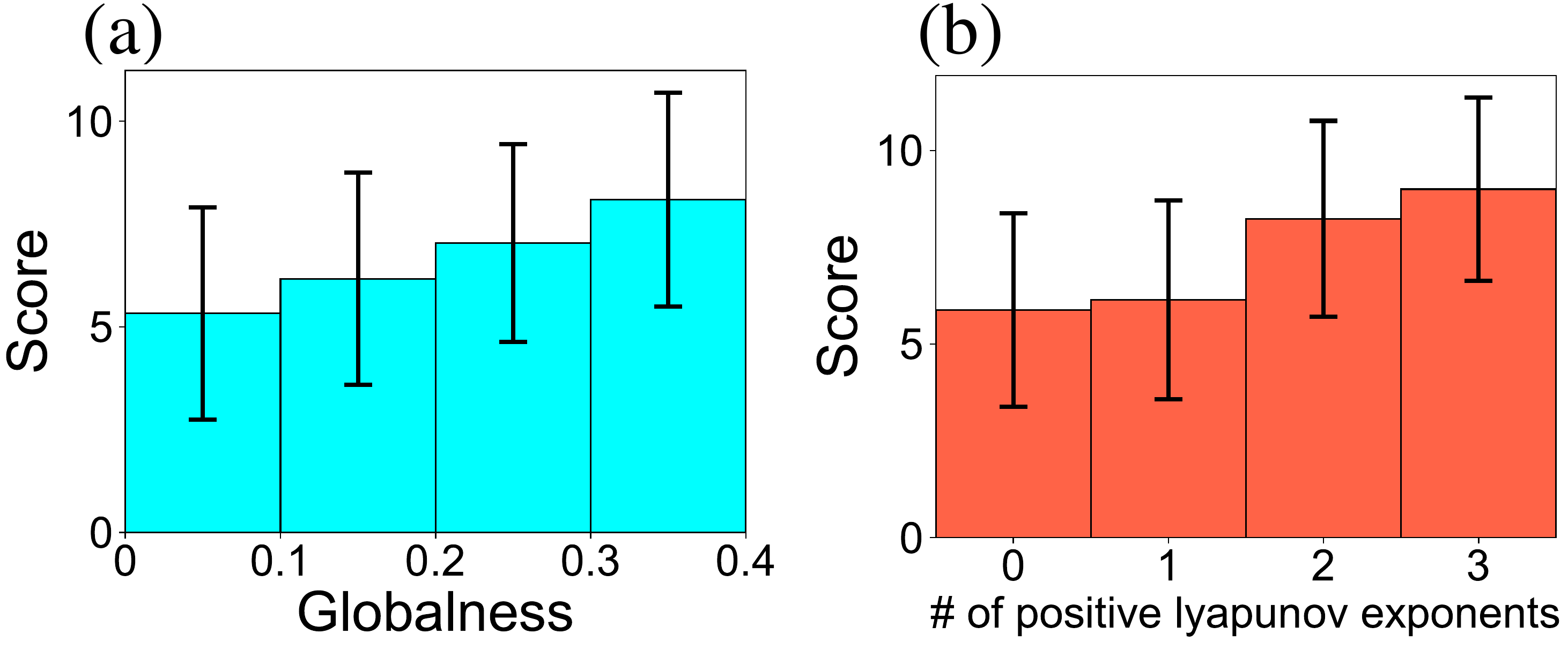}
    \caption{
        Fitness score versus the characteristics of the dynamics for fixed \( \theta_i = 0 \).
        (a) Scores as a function of the globalness of trajectories, as defined in Eq. \eqref{eq:gl}.
        (b) Scores as a function of the number of positive Lyapunov exponents.
        \( N = 60 \).
        }
        \label{Fig5}
\end{figure}

To examine if the global traveling of the orbit at \( \theta_i = 0 \) is relevant to adaptation, we computed the globalness of trajectories against the target patterns \( \{ X_m^k \} \), defined as 
\begin{align}
    G \equiv \frac{1}{K} \sum_k^K \left\{ < (\frac{1}{M} \sum_m^M x_m X_m^k)^2> -  < \frac{1}{M} \sum_m^M x_m X_m^k >^2 \right\}.
    \label{eq:gl}
\end{align}
In Fig. \ref{Fig5}, we plot the adaptation capacity against the globalness by sampling with 0.1 bin size and averaging the adaptation capacity for each bin.
As shown in Fig. \ref{Fig5}, the score (= adaptation capacity) monotonically increased with the globalness of trajectories.

As shown in Figs. \ref{Fig3} and \ref{Fig4}, such global traveling is supported by chaotic dynamics.
We computed the Lyapunov spectra of dynamics with fixed \( \theta_i = 0 \) and examined how they correlate with the adaptation capacity (fitness) of the system. 
Figure \ref{Fig5} shows the correlation in the fitness against the number of positive Lyapunov exponents, sampled over randomly chosen 500 GRNs for \( N=60 \).
The number of positive Lyapunov exponents gives the number of directions that tiny perturbations can be expanded.
Figure \ref{Fig5} suggests that the adaptation capacity increases with it.

\begin{figure}[tbp]
    \centering
    \includegraphics[width=0.7\linewidth]{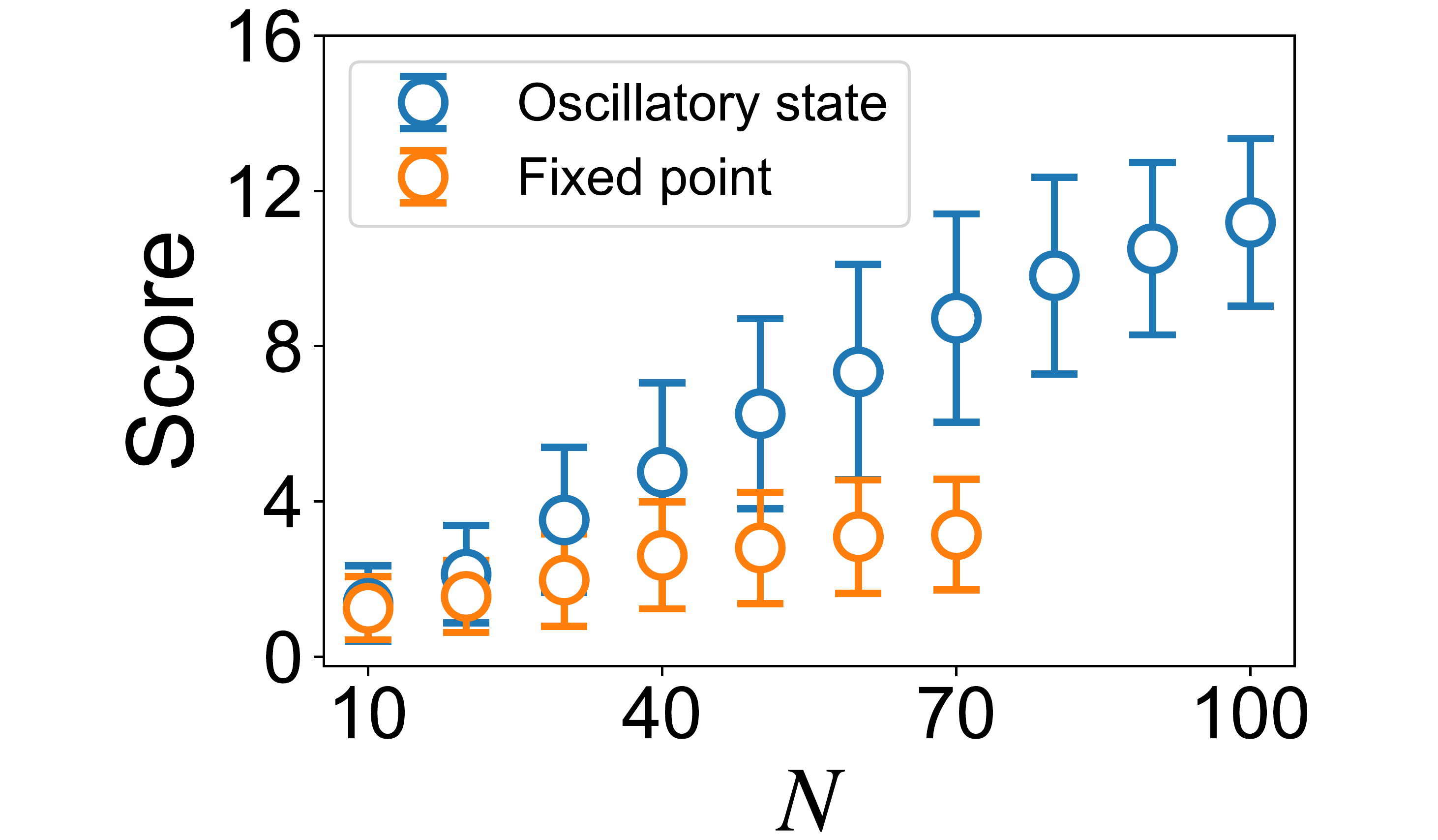}
    \caption{
        \( N \) dependency of adaptive environments in the randomly generated matrix \( J_{ij} \) for \( N = 10, 20, \dots, 90, 100 \).
        For each \( N \) we prepared 500 random gene regulatory matrix \( J_{ij} \) and averaged the score among them.
    }
    \label{Fig6}
\end{figure}

As the fraction of the network that gives oscillatory and (higher dimensional) chaotic dynamics is increased with the system size \cite{Sompolinsky1988-wx}, we expect the adaptation capacity is increased with the system size \( N \).
In Fig. \ref{Fig6}, we plot the average adaptation capacity of the networks with and without oscillatory dynamics for \( \theta_i = 0 \). (Note that for \( N>70 \) the networks without oscillatory dynamics cannot be sampled sufficiently).
As \( N \) is increased, more networks can adapt to almost all (\( = 2^M/2 \)) environmental conditions.

In this letter, we provide an adaptation mechanism based on fast oscillatory gene expression dynamics coupled with a slower epigenetic fixation process.
The chaotic oscillatory dynamics are relevant to the search for adapted states depending on the input, and once the adapted states are approached, a slower epigenetic modification process fixes such states.
As long as the searching by oscillatory (chaotic) dynamics sufficiently covers the state space, this adaptation mechanism works efficiently.
The degree of chaos or the region of the phase space traveled by orbits is correlated with the capacity of environments to which the cell can adapt.
As the number of genes (degrees of freedom) becomes large, the fraction of networks allowing for such dynamics is increased, supporting the generality of the proposed mechanism.

Previously, we reported that the interplay between gene expression oscillation and slow epigenetic feedback allows for robust cell differentiation needed for multicellular organisms \cite{Matsushita2020-fy,Matsushita2022-rx}.
As it is demonstrated herein that chaotic oscillatory dynamics and slower epigenetic fixation can allow for adaptation to multiple environments, this work may provide a path to understanding multicellular differentiation and unicellular adaptation coherently.
Our adaptation mechanism could be depicted as Waddington's epigenetic landscape that developmental biologists often adopt \cite{Waddington1957strategy}, wherein the initial state in a shallow valley travels over a large portion of phase space, as in the chaotic dynamics in our model, whereas with the slow epigenetic change, deep valleys are generated to which the cellular state is attracted to achieved adaptation. 

The proposed scheme requires neither attractors for adaptation in advance nor evolutionary optimization of the networks.
In this sense, it can support the generic and spontaneous adaptation of cells to unforeseen environmental conditions.
Thus far, there is no direct support for oscillatory expression dynamics in unicellular organisms despite some support in stem cells in multicellular organisms \cite{Palmeirim1997-kz, Huang2005-py, Chang2008-bj, Kobayashi2009-is, Zhang2019-wo}, as well as theoretical verifications \cite{Furusawa2012-er, Goto2013-dh, Koseska2013-hp}.
The expression dynamics are considerably noisy, and the experimental extraction of oscillatory components is challenging.
Notably, our mechanism works robustly under strong stochasticity.

The present model adopts a simple setup for oscillatory dynamics with on-off type dynamics and a slower fixation process.
Such on-off dynamics are ubiquitously adopted in biological and artificial neural networks.
The present scheme with an autonomous search for the desired state by chaotic dynamics and slower fixation can be generally applied to learning or optimization processes.
Here, as compared with random sampling adopted in simulated annealing \cite{Kirkpatrick1983-cu}, the chaotic dynamics do not need to sample the whole space, which will make the search more efficient \cite{Nozawa1994-yw, Tokuda1998-vm, Sinha1999-xu}.
We also note that the relevance of chaos or chaotic itinerancy to neural information processing has been discussed \cite{Skarda1987-gc, Kaneko1990-gi, Tsuda1992-tc, Tsuda2001-xn}, whereas \( \theta_i \) in the present model can regarded as inputs \cite{Kurikawa2013-wf}.
Moreover, in contrast to the Hebbian learning that requires the change in \( J_{ij} \) (i.e., \(N \times N\) elements), the present scheme requires the change only in \( \theta_i \) (i.e., \( N \) elements), which will be useful for the effectiveness in low-rank change in reservoir computation or echo-state networks \cite{Maass2002-jo, Yildiz2012-mq}.

\begin{acknowledgments}
The authors would like to thank Tetsuhiro S. Hatakeyama and Chikara Furusawa for their stimulating discussions.
This research was supported by a Grant-in-Aid for Scientific Research (A) 431 (20H00123) and from the Ministry of Education, Culture, Sports, Science, and Technology (MEXT) of Japan, and the Novo Nordisk Foundation.
\end{acknowledgments}


\bibliographystyle{apsrev4-2}

\begin{thebibliography}{46}%
\makeatletter
\providecommand \@ifxundefined [1]{%
 \@ifx{#1\undefined}
}%
\providecommand \@ifnum [1]{%
 \ifnum #1\expandafter \@firstoftwo
 \else \expandafter \@secondoftwo
 \fi
}%
\providecommand \@ifx [1]{%
 \ifx #1\expandafter \@firstoftwo
 \else \expandafter \@secondoftwo
 \fi
}%
\providecommand \natexlab [1]{#1}%
\providecommand \enquote  [1]{``#1''}%
\providecommand \bibnamefont  [1]{#1}%
\providecommand \bibfnamefont [1]{#1}%
\providecommand \citenamefont [1]{#1}%
\providecommand \href@noop [0]{\@secondoftwo}%
\providecommand \href [0]{\begingroup \@sanitize@url \@href}%
\providecommand \@href[1]{\@@startlink{#1}\@@href}%
\providecommand \@@href[1]{\endgroup#1\@@endlink}%
\providecommand \@sanitize@url [0]{\catcode `\\12\catcode `\$12\catcode
  `\&12\catcode `\#12\catcode `\^12\catcode `\_12\catcode `\%12\relax}%
\providecommand \@@startlink[1]{}%
\providecommand \@@endlink[0]{}%
\providecommand \url  [0]{\begingroup\@sanitize@url \@url }%
\providecommand \@url [1]{\endgroup\@href {#1}{\urlprefix }}%
\providecommand \urlprefix  [0]{URL }%
\providecommand \Eprint [0]{\href }%
\providecommand \doibase [0]{https://doi.org/}%
\providecommand \selectlanguage [0]{\@gobble}%
\providecommand \bibinfo  [0]{\@secondoftwo}%
\providecommand \bibfield  [0]{\@secondoftwo}%
\providecommand \translation [1]{[#1]}%
\providecommand \BibitemOpen [0]{}%
\providecommand \bibitemStop [0]{}%
\providecommand \bibitemNoStop [0]{.\EOS\space}%
\providecommand \EOS [0]{\spacefactor3000\relax}%
\providecommand \BibitemShut  [1]{\csname bibitem#1\endcsname}%
\let\auto@bib@innerbib\@empty
\bibitem [{\citenamefont {Alberts}(2017)}]{Alberts2017-sz}%
  \BibitemOpen
  \bibfield  {author} {\bibinfo {author} {\bibfnamefont {B.}~\bibnamefont
  {Alberts}},\ }\href@noop {} {\emph {\bibinfo {title} {Molecular biology of
  the cell}}}\ (\bibinfo  {publisher} {WW Norton \& Company},\ \bibinfo {year}
  {2017})\BibitemShut {NoStop}%
\bibitem [{\citenamefont {Braun}(2015)}]{Braun2015-wy}%
  \BibitemOpen
  \bibfield  {author} {\bibinfo {author} {\bibfnamefont {E.}~\bibnamefont
  {Braun}},\ }\href@noop {} {\bibfield  {journal} {\bibinfo  {journal} {Rep.
  Prog. Phys.}\ }\textbf {\bibinfo {volume} {78}},\ \bibinfo {pages} {036602}
  (\bibinfo {year} {2015})}\BibitemShut {NoStop}%
\bibitem [{\citenamefont {Kashiwagi}\ \emph {et~al.}(2006)\citenamefont
  {Kashiwagi}, \citenamefont {Urabe}, \citenamefont {Kaneko},\ and\
  \citenamefont {Yomo}}]{Kashiwagi2006-px}%
  \BibitemOpen
  \bibfield  {author} {\bibinfo {author} {\bibfnamefont {A.}~\bibnamefont
  {Kashiwagi}}, \bibinfo {author} {\bibfnamefont {I.}~\bibnamefont {Urabe}},
  \bibinfo {author} {\bibfnamefont {K.}~\bibnamefont {Kaneko}},\ and\ \bibinfo
  {author} {\bibfnamefont {T.}~\bibnamefont {Yomo}},\ }\href@noop {} {\bibfield
   {journal} {\bibinfo  {journal} {PLoS One}\ }\textbf {\bibinfo {volume}
  {1}},\ \bibinfo {pages} {e49} (\bibinfo {year} {2006})}\BibitemShut {NoStop}%
\bibitem [{\citenamefont {Ciechonska}\ \emph {et~al.}(2022)\citenamefont
  {Ciechonska}, \citenamefont {Sturrock}, \citenamefont {Grob}, \citenamefont
  {Larrouy-Maumus}, \citenamefont {Shahrezaei},\ and\ \citenamefont
  {Isalan}}]{Ciechonska2022-kl}%
  \BibitemOpen
  \bibfield  {author} {\bibinfo {author} {\bibfnamefont {M.}~\bibnamefont
  {Ciechonska}}, \bibinfo {author} {\bibfnamefont {M.}~\bibnamefont
  {Sturrock}}, \bibinfo {author} {\bibfnamefont {A.}~\bibnamefont {Grob}},
  \bibinfo {author} {\bibfnamefont {G.}~\bibnamefont {Larrouy-Maumus}},
  \bibinfo {author} {\bibfnamefont {V.}~\bibnamefont {Shahrezaei}},\ and\
  \bibinfo {author} {\bibfnamefont {M.}~\bibnamefont {Isalan}},\ }\href@noop {}
  {\bibfield  {journal} {\bibinfo  {journal} {PNAS Nexus}\ }\textbf {\bibinfo
  {volume} {1}},\ \bibinfo {pages} {gac069} (\bibinfo {year}
  {2022})}\BibitemShut {NoStop}%
\bibitem [{\citenamefont {Tsuru}\ \emph {et~al.}(2011)\citenamefont {Tsuru},
  \citenamefont {Yasuda}, \citenamefont {Murakami}, \citenamefont {Ushioda},
  \citenamefont {Kashiwagi}, \citenamefont {Suzuki}, \citenamefont {Mori},
  \citenamefont {Ying},\ and\ \citenamefont {Yomo}}]{Tsuru2011-mt}%
  \BibitemOpen
  \bibfield  {author} {\bibinfo {author} {\bibfnamefont {S.}~\bibnamefont
  {Tsuru}}, \bibinfo {author} {\bibfnamefont {N.}~\bibnamefont {Yasuda}},
  \bibinfo {author} {\bibfnamefont {Y.}~\bibnamefont {Murakami}}, \bibinfo
  {author} {\bibfnamefont {J.}~\bibnamefont {Ushioda}}, \bibinfo {author}
  {\bibfnamefont {A.}~\bibnamefont {Kashiwagi}}, \bibinfo {author}
  {\bibfnamefont {S.}~\bibnamefont {Suzuki}}, \bibinfo {author} {\bibfnamefont
  {K.}~\bibnamefont {Mori}}, \bibinfo {author} {\bibfnamefont {B.-W.}\
  \bibnamefont {Ying}},\ and\ \bibinfo {author} {\bibfnamefont
  {T.}~\bibnamefont {Yomo}},\ }\href@noop {} {\bibfield  {journal} {\bibinfo
  {journal} {Mol. Syst. Biol.}\ }\textbf {\bibinfo {volume} {7}},\ \bibinfo
  {pages} {493} (\bibinfo {year} {2011})}\BibitemShut {NoStop}%
\bibitem [{\citenamefont {Furusawa}\ and\ \citenamefont
  {Kaneko}(2008)}]{Furusawa2008-eo}%
  \BibitemOpen
  \bibfield  {author} {\bibinfo {author} {\bibfnamefont {C.}~\bibnamefont
  {Furusawa}}\ and\ \bibinfo {author} {\bibfnamefont {K.}~\bibnamefont
  {Kaneko}},\ }\href@noop {} {\bibfield  {journal} {\bibinfo  {journal} {PLoS
  Comput. Biol.}\ }\textbf {\bibinfo {volume} {4}},\ \bibinfo {pages} {e3}
  (\bibinfo {year} {2008})}\BibitemShut {NoStop}%
\bibitem [{\citenamefont {Furusawa}\ and\ \citenamefont
  {Kaneko}(2013)}]{Furusawa2013-pc}%
  \BibitemOpen
  \bibfield  {author} {\bibinfo {author} {\bibfnamefont {C.}~\bibnamefont
  {Furusawa}}\ and\ \bibinfo {author} {\bibfnamefont {K.}~\bibnamefont
  {Kaneko}},\ }\href@noop {} {\bibfield  {journal} {\bibinfo  {journal} {PLoS
  One}\ }\textbf {\bibinfo {volume} {8}},\ \bibinfo {pages} {e61251} (\bibinfo
  {year} {2013})}\BibitemShut {NoStop}%
\bibitem [{\citenamefont {Leibnitz}\ and\ \citenamefont
  {Murata}(2010)}]{Leibnitz2010-nf}%
  \BibitemOpen
  \bibfield  {author} {\bibinfo {author} {\bibfnamefont {K.}~\bibnamefont
  {Leibnitz}}\ and\ \bibinfo {author} {\bibfnamefont {M.}~\bibnamefont
  {Murata}},\ }\href@noop {} {\bibfield  {journal} {\bibinfo  {journal} {IEEE
  Netw.}\ }\textbf {\bibinfo {volume} {24}},\ \bibinfo {pages} {14} (\bibinfo
  {year} {2010})}\BibitemShut {NoStop}%
\bibitem [{\citenamefont {Koizumi}\ \emph {et~al.}(2010)\citenamefont
  {Koizumi}, \citenamefont {Miyamura}, \citenamefont {Arakawa}, \citenamefont
  {Oki}, \citenamefont {Shiomoto},\ and\ \citenamefont
  {Murata}}]{Koizumi2010-qw}%
  \BibitemOpen
  \bibfield  {author} {\bibinfo {author} {\bibfnamefont {Y.}~\bibnamefont
  {Koizumi}}, \bibinfo {author} {\bibfnamefont {T.}~\bibnamefont {Miyamura}},
  \bibinfo {author} {\bibfnamefont {S.}~\bibnamefont {Arakawa}}, \bibinfo
  {author} {\bibfnamefont {E.}~\bibnamefont {Oki}}, \bibinfo {author}
  {\bibfnamefont {K.}~\bibnamefont {Shiomoto}},\ and\ \bibinfo {author}
  {\bibfnamefont {M.}~\bibnamefont {Murata}},\ }\href@noop {} {\bibfield
  {journal} {\bibinfo  {journal} {J. Lightwave Technol.}\ }\textbf {\bibinfo
  {volume} {28}},\ \bibinfo {pages} {1720} (\bibinfo {year}
  {2010})}\BibitemShut {NoStop}%
\bibitem [{\citenamefont {Gombar}\ \emph {et~al.}(2014)\citenamefont {Gombar},
  \citenamefont {MacCarthy},\ and\ \citenamefont {Bergman}}]{Gombar2014-rv}%
  \BibitemOpen
  \bibfield  {author} {\bibinfo {author} {\bibfnamefont {S.}~\bibnamefont
  {Gombar}}, \bibinfo {author} {\bibfnamefont {T.}~\bibnamefont {MacCarthy}},\
  and\ \bibinfo {author} {\bibfnamefont {A.}~\bibnamefont {Bergman}},\
  }\href@noop {} {\bibfield  {journal} {\bibinfo  {journal} {PLoS Comput.
  Biol.}\ }\textbf {\bibinfo {volume} {10}},\ \bibinfo {pages} {e1003450}
  (\bibinfo {year} {2014})}\BibitemShut {NoStop}%
\bibitem [{\citenamefont {Matsushita}\ and\ \citenamefont
  {Kaneko}(2020)}]{Matsushita2020-fy}%
  \BibitemOpen
  \bibfield  {author} {\bibinfo {author} {\bibfnamefont {Y.}~\bibnamefont
  {Matsushita}}\ and\ \bibinfo {author} {\bibfnamefont {K.}~\bibnamefont
  {Kaneko}},\ }\href@noop {} {\bibfield  {journal} {\bibinfo  {journal} {Phys.
  Rev. Research}\ }\textbf {\bibinfo {volume} {2}},\ \bibinfo {pages} {023083}
  (\bibinfo {year} {2020})}\BibitemShut {NoStop}%
\bibitem [{\citenamefont {Matsushita}\ \emph {et~al.}(2022)\citenamefont
  {Matsushita}, \citenamefont {Hatakeyama},\ and\ \citenamefont
  {Kaneko}}]{Matsushita2022-rx}%
  \BibitemOpen
  \bibfield  {author} {\bibinfo {author} {\bibfnamefont {Y.}~\bibnamefont
  {Matsushita}}, \bibinfo {author} {\bibfnamefont {T.~S.}\ \bibnamefont
  {Hatakeyama}},\ and\ \bibinfo {author} {\bibfnamefont {K.}~\bibnamefont
  {Kaneko}},\ }\href@noop {} {\bibfield  {journal} {\bibinfo  {journal} {Phys.
  Rev. Research}\ }\textbf {\bibinfo {volume} {4}},\ \bibinfo {pages} {L022008}
  (\bibinfo {year} {2022})}\BibitemShut {NoStop}%
\bibitem [{\citenamefont {Bird}(2007)}]{Bird2007-tw}%
  \BibitemOpen
  \bibfield  {author} {\bibinfo {author} {\bibfnamefont {A.}~\bibnamefont
  {Bird}},\ }\href@noop {} {\bibfield  {journal} {\bibinfo  {journal} {Nature}\
  }\textbf {\bibinfo {volume} {447}},\ \bibinfo {pages} {396} (\bibinfo {year}
  {2007})}\BibitemShut {NoStop}%
\bibitem [{\citenamefont {Cortini}\ \emph {et~al.}(2016)\citenamefont
  {Cortini}, \citenamefont {Barbi}, \citenamefont {Car{\'e}}, \citenamefont
  {Lavelle}, \citenamefont {Lesne}, \citenamefont {Mozziconacci},\ and\
  \citenamefont {Victor}}]{Cortini2016-vu}%
  \BibitemOpen
  \bibfield  {author} {\bibinfo {author} {\bibfnamefont {R.}~\bibnamefont
  {Cortini}}, \bibinfo {author} {\bibfnamefont {M.}~\bibnamefont {Barbi}},
  \bibinfo {author} {\bibfnamefont {B.~R.}\ \bibnamefont {Car{\'e}}}, \bibinfo
  {author} {\bibfnamefont {C.}~\bibnamefont {Lavelle}}, \bibinfo {author}
  {\bibfnamefont {A.}~\bibnamefont {Lesne}}, \bibinfo {author} {\bibfnamefont
  {J.}~\bibnamefont {Mozziconacci}},\ and\ \bibinfo {author} {\bibfnamefont
  {J.-M.}\ \bibnamefont {Victor}},\ }\href@noop {} {\bibfield  {journal}
  {\bibinfo  {journal} {Rev. Mod. Phys.}\ }\textbf {\bibinfo {volume} {88}},\
  \bibinfo {pages} {025002} (\bibinfo {year} {2016})}\BibitemShut {NoStop}%
\bibitem [{\citenamefont {Hihara}\ \emph {et~al.}(2012)\citenamefont {Hihara},
  \citenamefont {Pack}, \citenamefont {Kaizu}, \citenamefont {Tani},
  \citenamefont {Hanafusa}, \citenamefont {Nozaki}, \citenamefont {Takemoto},
  \citenamefont {Yoshimi}, \citenamefont {Yokota}, \citenamefont {Imamoto},
  \citenamefont {Sako}, \citenamefont {Kinjo}, \citenamefont {Takahashi},
  \citenamefont {Nagai},\ and\ \citenamefont {Maeshima}}]{Hihara2012-ga}%
  \BibitemOpen
  \bibfield  {author} {\bibinfo {author} {\bibfnamefont {S.}~\bibnamefont
  {Hihara}}, \bibinfo {author} {\bibfnamefont {C.-G.}\ \bibnamefont {Pack}},
  \bibinfo {author} {\bibfnamefont {K.}~\bibnamefont {Kaizu}}, \bibinfo
  {author} {\bibfnamefont {T.}~\bibnamefont {Tani}}, \bibinfo {author}
  {\bibfnamefont {T.}~\bibnamefont {Hanafusa}}, \bibinfo {author}
  {\bibfnamefont {T.}~\bibnamefont {Nozaki}}, \bibinfo {author} {\bibfnamefont
  {S.}~\bibnamefont {Takemoto}}, \bibinfo {author} {\bibfnamefont
  {T.}~\bibnamefont {Yoshimi}}, \bibinfo {author} {\bibfnamefont
  {H.}~\bibnamefont {Yokota}}, \bibinfo {author} {\bibfnamefont
  {N.}~\bibnamefont {Imamoto}}, \bibinfo {author} {\bibfnamefont
  {Y.}~\bibnamefont {Sako}}, \bibinfo {author} {\bibfnamefont {M.}~\bibnamefont
  {Kinjo}}, \bibinfo {author} {\bibfnamefont {K.}~\bibnamefont {Takahashi}},
  \bibinfo {author} {\bibfnamefont {T.}~\bibnamefont {Nagai}},\ and\ \bibinfo
  {author} {\bibfnamefont {K.}~\bibnamefont {Maeshima}},\ }\href@noop {}
  {\bibfield  {journal} {\bibinfo  {journal} {Cell Rep.}\ }\textbf {\bibinfo
  {volume} {2}},\ \bibinfo {pages} {1645} (\bibinfo {year} {2012})}\BibitemShut
  {NoStop}%
\bibitem [{\citenamefont {Tripathi}\ and\ \citenamefont
  {Menon}(2019)}]{Tripathi2019-de}%
  \BibitemOpen
  \bibfield  {author} {\bibinfo {author} {\bibfnamefont {K.}~\bibnamefont
  {Tripathi}}\ and\ \bibinfo {author} {\bibfnamefont {G.~I.}\ \bibnamefont
  {Menon}},\ }\href@noop {} {\bibfield  {journal} {\bibinfo  {journal} {Phys.
  Rev. X}\ }\textbf {\bibinfo {volume} {9}},\ \bibinfo {pages} {041020}
  (\bibinfo {year} {2019})}\BibitemShut {NoStop}%
\bibitem [{Note1()}]{Note1}%
  \BibitemOpen
  \bibinfo {note} {Here, we describe the model in terms of genes; however, it
  can be generalized to introducing units with a slow feedback fixation
  process.}\BibitemShut {Stop}%
\bibitem [{\citenamefont {Mjolsness}\ \emph {et~al.}(1991)\citenamefont
  {Mjolsness}, \citenamefont {Sharp},\ and\ \citenamefont
  {Reinitz}}]{Mjolsness1991-vr}%
  \BibitemOpen
  \bibfield  {author} {\bibinfo {author} {\bibfnamefont {E.}~\bibnamefont
  {Mjolsness}}, \bibinfo {author} {\bibfnamefont {D.~H.}\ \bibnamefont
  {Sharp}},\ and\ \bibinfo {author} {\bibfnamefont {J.}~\bibnamefont
  {Reinitz}},\ }\href@noop {} {\bibfield  {journal} {\bibinfo  {journal} {J.
  Theor. Biol.}\ } (\bibinfo {year} {1991})}\BibitemShut {NoStop}%
\bibitem [{\citenamefont {Salazar-Ciudad}\ \emph {et~al.}(2000)\citenamefont
  {Salazar-Ciudad}, \citenamefont {Garcia-Fern{\'a}ndez},\ and\ \citenamefont
  {Sol{\'e}}}]{Salazar-Ciudad2000-ov}%
  \BibitemOpen
  \bibfield  {author} {\bibinfo {author} {\bibfnamefont {I.}~\bibnamefont
  {Salazar-Ciudad}}, \bibinfo {author} {\bibfnamefont {J.}~\bibnamefont
  {Garcia-Fern{\'a}ndez}},\ and\ \bibinfo {author} {\bibfnamefont {R.~V.}\
  \bibnamefont {Sol{\'e}}},\ }\href@noop {} {\bibfield  {journal} {\bibinfo
  {journal} {J. Theor. Biol.}\ }\textbf {\bibinfo {volume} {205}},\ \bibinfo
  {pages} {587} (\bibinfo {year} {2000})}\BibitemShut {NoStop}%
\bibitem [{\citenamefont {Salazar-Ciudad}\ \emph {et~al.}(2001)\citenamefont
  {Salazar-Ciudad}, \citenamefont {Newman},\ and\ \citenamefont
  {Sol{\'e}}}]{Salazar-Ciudad2001-qt}%
  \BibitemOpen
  \bibfield  {author} {\bibinfo {author} {\bibfnamefont {I.}~\bibnamefont
  {Salazar-Ciudad}}, \bibinfo {author} {\bibfnamefont {S.~A.}\ \bibnamefont
  {Newman}},\ and\ \bibinfo {author} {\bibfnamefont {R.~V.}\ \bibnamefont
  {Sol{\'e}}},\ }\href@noop {} {\bibfield  {journal} {\bibinfo  {journal}
  {Evol. Dev.}\ }\textbf {\bibinfo {volume} {3}},\ \bibinfo {pages} {84}
  (\bibinfo {year} {2001})}\BibitemShut {NoStop}%
\bibitem [{\citenamefont {Miyamoto}\ \emph {et~al.}(2015)\citenamefont
  {Miyamoto}, \citenamefont {Furusawa},\ and\ \citenamefont
  {Kaneko}}]{Miyamoto2015-ux}%
  \BibitemOpen
  \bibfield  {author} {\bibinfo {author} {\bibfnamefont {T.}~\bibnamefont
  {Miyamoto}}, \bibinfo {author} {\bibfnamefont {C.}~\bibnamefont {Furusawa}},\
  and\ \bibinfo {author} {\bibfnamefont {K.}~\bibnamefont {Kaneko}},\
  }\href@noop {} {\bibfield  {journal} {\bibinfo  {journal} {PLoS Comput.
  Biol.}\ } (\bibinfo {year} {2015})}\BibitemShut {NoStop}%
\bibitem [{\citenamefont {Huang}\ \emph {et~al.}(2020)\citenamefont {Huang},
  \citenamefont {Lu}, \citenamefont {Galbraith}, \citenamefont {Levine},
  \citenamefont {Onuchic},\ and\ \citenamefont {Jia}}]{Huang2020-yk}%
  \BibitemOpen
  \bibfield  {author} {\bibinfo {author} {\bibfnamefont {B.}~\bibnamefont
  {Huang}}, \bibinfo {author} {\bibfnamefont {M.}~\bibnamefont {Lu}}, \bibinfo
  {author} {\bibfnamefont {M.}~\bibnamefont {Galbraith}}, \bibinfo {author}
  {\bibfnamefont {H.}~\bibnamefont {Levine}}, \bibinfo {author} {\bibfnamefont
  {J.~N.}\ \bibnamefont {Onuchic}},\ and\ \bibinfo {author} {\bibfnamefont
  {D.}~\bibnamefont {Jia}},\ }\href@noop {} {\bibfield  {journal} {\bibinfo
  {journal} {J. R. Soc. Interface}\ }\textbf {\bibinfo {volume} {17}},\
  \bibinfo {pages} {20200500} (\bibinfo {year} {2020})}\BibitemShut {NoStop}%
\bibitem [{\citenamefont {Schreiber}\ and\ \citenamefont
  {Bernstein}(2002)}]{Schreiber2002-gm}%
  \BibitemOpen
  \bibfield  {author} {\bibinfo {author} {\bibfnamefont {S.~L.}\ \bibnamefont
  {Schreiber}}\ and\ \bibinfo {author} {\bibfnamefont {B.~E.}\ \bibnamefont
  {Bernstein}},\ }\href@noop {} {\bibfield  {journal} {\bibinfo  {journal}
  {Cell}\ }\textbf {\bibinfo {volume} {111}},\ \bibinfo {pages} {771} (\bibinfo
  {year} {2002})}\BibitemShut {NoStop}%
\bibitem [{\citenamefont {Dodd}\ \emph {et~al.}(2007)\citenamefont {Dodd},
  \citenamefont {Micheelsen}, \citenamefont {Sneppen},\ and\ \citenamefont
  {Thon}}]{Dodd2007-kz}%
  \BibitemOpen
  \bibfield  {author} {\bibinfo {author} {\bibfnamefont {I.~B.}\ \bibnamefont
  {Dodd}}, \bibinfo {author} {\bibfnamefont {M.~A.}\ \bibnamefont
  {Micheelsen}}, \bibinfo {author} {\bibfnamefont {K.}~\bibnamefont
  {Sneppen}},\ and\ \bibinfo {author} {\bibfnamefont {G.}~\bibnamefont
  {Thon}},\ }\href@noop {} {\bibfield  {journal} {\bibinfo  {journal} {Cell}\
  }\textbf {\bibinfo {volume} {129}},\ \bibinfo {pages} {813} (\bibinfo {year}
  {2007})}\BibitemShut {NoStop}%
\bibitem [{\citenamefont {Sneppen}\ \emph {et~al.}(2008)\citenamefont
  {Sneppen}, \citenamefont {Micheelsen},\ and\ \citenamefont
  {Dodd}}]{Sneppen2008-ui}%
  \BibitemOpen
  \bibfield  {author} {\bibinfo {author} {\bibfnamefont {K.}~\bibnamefont
  {Sneppen}}, \bibinfo {author} {\bibfnamefont {M.~A.}\ \bibnamefont
  {Micheelsen}},\ and\ \bibinfo {author} {\bibfnamefont {I.~B.}\ \bibnamefont
  {Dodd}},\ }\href@noop {} {\bibfield  {journal} {\bibinfo  {journal} {Mol.
  Syst. Biol.}\ }\textbf {\bibinfo {volume} {4}},\ \bibinfo {pages} {182}
  (\bibinfo {year} {2008})}\BibitemShut {NoStop}%
\bibitem [{\citenamefont {Sompolinsky}\ \emph {et~al.}(1988)\citenamefont
  {Sompolinsky}, \citenamefont {Crisanti},\ and\ \citenamefont
  {Sommers}}]{Sompolinsky1988-wx}%
  \BibitemOpen
  \bibfield  {author} {\bibinfo {author} {\bibfnamefont {H.}~\bibnamefont
  {Sompolinsky}}, \bibinfo {author} {\bibfnamefont {A.}~\bibnamefont
  {Crisanti}},\ and\ \bibinfo {author} {\bibfnamefont {H.~J.}\ \bibnamefont
  {Sommers}},\ }\href@noop {} {\bibfield  {journal} {\bibinfo  {journal} {Phys.
  Rev. Lett.}\ }\textbf {\bibinfo {volume} {61}},\ \bibinfo {pages} {259}
  (\bibinfo {year} {1988})}\BibitemShut {NoStop}%
\bibitem [{\citenamefont {Waddington}(1957)}]{Waddington1957strategy}%
  \BibitemOpen
  \bibfield  {author} {\bibinfo {author} {\bibfnamefont {C.}~\bibnamefont
  {Waddington}},\ }\href@noop {} {\emph {\bibinfo {title} {The Strategy of the
  Genes}}}\ (\bibinfo  {publisher} {George Allen \& Unwin},\ \bibinfo {year}
  {1957})\BibitemShut {NoStop}%
\bibitem [{\citenamefont {Palmeirim}\ \emph {et~al.}(1997)\citenamefont
  {Palmeirim}, \citenamefont {Henrique}, \citenamefont {Ish-Horowicz},\ and\
  \citenamefont {Pourqui{\'e}}}]{Palmeirim1997-kz}%
  \BibitemOpen
  \bibfield  {author} {\bibinfo {author} {\bibfnamefont {I.}~\bibnamefont
  {Palmeirim}}, \bibinfo {author} {\bibfnamefont {D.}~\bibnamefont {Henrique}},
  \bibinfo {author} {\bibfnamefont {D.}~\bibnamefont {Ish-Horowicz}},\ and\
  \bibinfo {author} {\bibfnamefont {O.}~\bibnamefont {Pourqui{\'e}}},\
  }\href@noop {} {\bibfield  {journal} {\bibinfo  {journal} {Cell}\ }\textbf
  {\bibinfo {volume} {91}},\ \bibinfo {pages} {639} (\bibinfo {year}
  {1997})}\BibitemShut {NoStop}%
\bibitem [{\citenamefont {Huang}\ \emph {et~al.}(2005)\citenamefont {Huang},
  \citenamefont {Eichler}, \citenamefont {Bar-Yam},\ and\ \citenamefont
  {Ingber}}]{Huang2005-py}%
  \BibitemOpen
  \bibfield  {author} {\bibinfo {author} {\bibfnamefont {S.}~\bibnamefont
  {Huang}}, \bibinfo {author} {\bibfnamefont {G.}~\bibnamefont {Eichler}},
  \bibinfo {author} {\bibfnamefont {Y.}~\bibnamefont {Bar-Yam}},\ and\ \bibinfo
  {author} {\bibfnamefont {D.~E.}\ \bibnamefont {Ingber}},\ }\href
  {https://doi.org/10.1103/PhysRevLett.94.128701} {\bibfield  {journal}
  {\bibinfo  {journal} {Phys. Rev. Lett.}\ }\textbf {\bibinfo {volume} {94}},\
  \bibinfo {pages} {128701} (\bibinfo {year} {2005})}\BibitemShut {NoStop}%
\bibitem [{\citenamefont {Chang}\ \emph {et~al.}(2008)\citenamefont {Chang},
  \citenamefont {Hemberg}, \citenamefont {Barahona}, \citenamefont {Ingber},\
  and\ \citenamefont {Huang}}]{Chang2008-bj}%
  \BibitemOpen
  \bibfield  {author} {\bibinfo {author} {\bibfnamefont {H.~H.}\ \bibnamefont
  {Chang}}, \bibinfo {author} {\bibfnamefont {M.}~\bibnamefont {Hemberg}},
  \bibinfo {author} {\bibfnamefont {M.}~\bibnamefont {Barahona}}, \bibinfo
  {author} {\bibfnamefont {D.~E.}\ \bibnamefont {Ingber}},\ and\ \bibinfo
  {author} {\bibfnamefont {S.}~\bibnamefont {Huang}},\ }\href@noop {}
  {\bibfield  {journal} {\bibinfo  {journal} {Nature}\ }\textbf {\bibinfo
  {volume} {453}},\ \bibinfo {pages} {544} (\bibinfo {year}
  {2008})}\BibitemShut {NoStop}%
\bibitem [{\citenamefont {Kobayashi}\ \emph {et~al.}(2009)\citenamefont
  {Kobayashi}, \citenamefont {Mizuno}, \citenamefont {Imayoshi}, \citenamefont
  {Furusawa}, \citenamefont {Shirahige},\ and\ \citenamefont
  {Kageyama}}]{Kobayashi2009-is}%
  \BibitemOpen
  \bibfield  {author} {\bibinfo {author} {\bibfnamefont {T.}~\bibnamefont
  {Kobayashi}}, \bibinfo {author} {\bibfnamefont {H.}~\bibnamefont {Mizuno}},
  \bibinfo {author} {\bibfnamefont {I.}~\bibnamefont {Imayoshi}}, \bibinfo
  {author} {\bibfnamefont {C.}~\bibnamefont {Furusawa}}, \bibinfo {author}
  {\bibfnamefont {K.}~\bibnamefont {Shirahige}},\ and\ \bibinfo {author}
  {\bibfnamefont {R.}~\bibnamefont {Kageyama}},\ }\href@noop {} {\bibfield
  {journal} {\bibinfo  {journal} {Genes Dev.}\ }\textbf {\bibinfo {volume}
  {23}},\ \bibinfo {pages} {1870} (\bibinfo {year} {2009})}\BibitemShut
  {NoStop}%
\bibitem [{\citenamefont {Zhang}\ \emph {et~al.}(2019)\citenamefont {Zhang},
  \citenamefont {Nie},\ and\ \citenamefont {Zhou}}]{Zhang2019-wo}%
  \BibitemOpen
  \bibfield  {author} {\bibinfo {author} {\bibfnamefont {J.}~\bibnamefont
  {Zhang}}, \bibinfo {author} {\bibfnamefont {Q.}~\bibnamefont {Nie}},\ and\
  \bibinfo {author} {\bibfnamefont {T.}~\bibnamefont {Zhou}},\ }\href@noop {}
  {\bibfield  {journal} {\bibinfo  {journal} {Front. Genet.}\ }\textbf
  {\bibinfo {volume} {10}},\ \bibinfo {pages} {1280} (\bibinfo {year}
  {2019})}\BibitemShut {NoStop}%
\bibitem [{\citenamefont {Furusawa}\ and\ \citenamefont
  {Kaneko}(2012)}]{Furusawa2012-er}%
  \BibitemOpen
  \bibfield  {author} {\bibinfo {author} {\bibfnamefont {C.}~\bibnamefont
  {Furusawa}}\ and\ \bibinfo {author} {\bibfnamefont {K.}~\bibnamefont
  {Kaneko}},\ }\href@noop {} {\bibfield  {journal} {\bibinfo  {journal}
  {Science}\ }\textbf {\bibinfo {volume} {338}},\ \bibinfo {pages} {215}
  (\bibinfo {year} {2012})}\BibitemShut {NoStop}%
\bibitem [{\citenamefont {Goto}\ and\ \citenamefont
  {Kaneko}(2013)}]{Goto2013-dh}%
  \BibitemOpen
  \bibfield  {author} {\bibinfo {author} {\bibfnamefont {Y.}~\bibnamefont
  {Goto}}\ and\ \bibinfo {author} {\bibfnamefont {K.}~\bibnamefont {Kaneko}},\
  }\href@noop {} {\bibfield  {journal} {\bibinfo  {journal} {Phys. Rev. E Stat.
  Nonlin. Soft Matter Phys.}\ }\textbf {\bibinfo {volume} {88}},\ \bibinfo
  {pages} {032718} (\bibinfo {year} {2013})}\BibitemShut {NoStop}%
\bibitem [{\citenamefont {Koseska}\ \emph {et~al.}(2013)\citenamefont
  {Koseska}, \citenamefont {Volkov},\ and\ \citenamefont
  {Kurths}}]{Koseska2013-hp}%
  \BibitemOpen
  \bibfield  {author} {\bibinfo {author} {\bibfnamefont {A.}~\bibnamefont
  {Koseska}}, \bibinfo {author} {\bibfnamefont {E.}~\bibnamefont {Volkov}},\
  and\ \bibinfo {author} {\bibfnamefont {J.}~\bibnamefont {Kurths}},\
  }\href@noop {} {\bibfield  {journal} {\bibinfo  {journal} {Phys. Rev. Lett.}\
  }\textbf {\bibinfo {volume} {111}},\ \bibinfo {pages} {024103} (\bibinfo
  {year} {2013})}\BibitemShut {NoStop}%
\bibitem [{\citenamefont {Kirkpatrick}\ \emph {et~al.}(1983)\citenamefont
  {Kirkpatrick}, \citenamefont {Gelatt},\ and\ \citenamefont
  {Vecchi}}]{Kirkpatrick1983-cu}%
  \BibitemOpen
  \bibfield  {author} {\bibinfo {author} {\bibfnamefont {S.}~\bibnamefont
  {Kirkpatrick}}, \bibinfo {author} {\bibfnamefont {C.~D.}\ \bibnamefont
  {Gelatt}, \bibfnamefont {Jr}},\ and\ \bibinfo {author} {\bibfnamefont
  {M.~P.}\ \bibnamefont {Vecchi}},\ }\href@noop {} {\bibfield  {journal}
  {\bibinfo  {journal} {Science}\ }\textbf {\bibinfo {volume} {220}},\ \bibinfo
  {pages} {671} (\bibinfo {year} {1983})}\BibitemShut {NoStop}%
\bibitem [{\citenamefont {Nozawa}(1994)}]{Nozawa1994-yw}%
  \BibitemOpen
  \bibfield  {author} {\bibinfo {author} {\bibfnamefont {H.}~\bibnamefont
  {Nozawa}},\ }\href@noop {} {\bibfield  {journal} {\bibinfo  {journal}
  {Physica D}\ }\textbf {\bibinfo {volume} {75}},\ \bibinfo {pages} {179}
  (\bibinfo {year} {1994})}\BibitemShut {NoStop}%
\bibitem [{\citenamefont {Tokuda}\ \emph {et~al.}(1998)\citenamefont {Tokuda},
  \citenamefont {Aihara},\ and\ \citenamefont {Nagashima}}]{Tokuda1998-vm}%
  \BibitemOpen
  \bibfield  {author} {\bibinfo {author} {\bibfnamefont {I.}~\bibnamefont
  {Tokuda}}, \bibinfo {author} {\bibfnamefont {K.}~\bibnamefont {Aihara}},\
  and\ \bibinfo {author} {\bibfnamefont {T.}~\bibnamefont {Nagashima}},\
  }\href@noop {} {\bibfield  {journal} {\bibinfo  {journal} {Phys. Rev. E}\
  }\textbf {\bibinfo {volume} {58}},\ \bibinfo {pages} {5157} (\bibinfo {year}
  {1998})}\BibitemShut {NoStop}%
\bibitem [{\citenamefont {Sinha}\ and\ \citenamefont
  {Ditto}(1999)}]{Sinha1999-xu}%
  \BibitemOpen
  \bibfield  {author} {\bibinfo {author} {\bibfnamefont {S.}~\bibnamefont
  {Sinha}}\ and\ \bibinfo {author} {\bibfnamefont {W.~L.}\ \bibnamefont
  {Ditto}},\ }\href@noop {} {\bibfield  {journal} {\bibinfo  {journal} {Phys.
  Rev. E Stat. Phys. Plasmas Fluids Relat. Interdiscip. Topics}\ }\textbf
  {\bibinfo {volume} {60}},\ \bibinfo {pages} {363} (\bibinfo {year}
  {1999})}\BibitemShut {NoStop}%
\bibitem [{\citenamefont {Skarda}\ and\ \citenamefont
  {Freeman}(1987)}]{Skarda1987-gc}%
  \BibitemOpen
  \bibfield  {author} {\bibinfo {author} {\bibfnamefont {C.~A.}\ \bibnamefont
  {Skarda}}\ and\ \bibinfo {author} {\bibfnamefont {W.~J.}\ \bibnamefont
  {Freeman}},\ }\href@noop {} {\bibfield  {journal} {\bibinfo  {journal}
  {Behav. Brain Sci.}\ }\textbf {\bibinfo {volume} {10}},\ \bibinfo {pages}
  {161} (\bibinfo {year} {1987})}\BibitemShut {NoStop}%
\bibitem [{\citenamefont {Kaneko}(1990)}]{Kaneko1990-gi}%
  \BibitemOpen
  \bibfield  {author} {\bibinfo {author} {\bibfnamefont {K.}~\bibnamefont
  {Kaneko}},\ }\href@noop {} {\bibfield  {journal} {\bibinfo  {journal}
  {Physica D}\ }\textbf {\bibinfo {volume} {41}},\ \bibinfo {pages} {137}
  (\bibinfo {year} {1990})}\BibitemShut {NoStop}%
\bibitem [{\citenamefont {Tsuda}(1992)}]{Tsuda1992-tc}%
  \BibitemOpen
  \bibfield  {author} {\bibinfo {author} {\bibfnamefont {I.}~\bibnamefont
  {Tsuda}},\ }\href@noop {} {\bibfield  {journal} {\bibinfo  {journal} {Neural
  Netw.}\ }\textbf {\bibinfo {volume} {5}},\ \bibinfo {pages} {313} (\bibinfo
  {year} {1992})}\BibitemShut {NoStop}%
\bibitem [{\citenamefont {Tsuda}(2001)}]{Tsuda2001-xn}%
  \BibitemOpen
  \bibfield  {author} {\bibinfo {author} {\bibfnamefont {I.}~\bibnamefont
  {Tsuda}},\ }\href@noop {} {\bibfield  {journal} {\bibinfo  {journal} {Behav.
  Brain Sci.}\ }\textbf {\bibinfo {volume} {24}},\ \bibinfo {pages} {793}
  (\bibinfo {year} {2001})}\BibitemShut {NoStop}%
\bibitem [{\citenamefont {Kurikawa}\ and\ \citenamefont
  {Kaneko}(2013)}]{Kurikawa2013-wf}%
  \BibitemOpen
  \bibfield  {author} {\bibinfo {author} {\bibfnamefont {T.}~\bibnamefont
  {Kurikawa}}\ and\ \bibinfo {author} {\bibfnamefont {K.}~\bibnamefont
  {Kaneko}},\ }\href@noop {} {\bibfield  {journal} {\bibinfo  {journal} {PLoS
  Comput. Biol.}\ }\textbf {\bibinfo {volume} {9}},\ \bibinfo {pages}
  {e1002943} (\bibinfo {year} {2013})}\BibitemShut {NoStop}%
\bibitem [{\citenamefont {Maass}\ \emph {et~al.}(2002)\citenamefont {Maass},
  \citenamefont {Natschl{\"a}ger},\ and\ \citenamefont
  {Markram}}]{Maass2002-jo}%
  \BibitemOpen
  \bibfield  {author} {\bibinfo {author} {\bibfnamefont {W.}~\bibnamefont
  {Maass}}, \bibinfo {author} {\bibfnamefont {T.}~\bibnamefont
  {Natschl{\"a}ger}},\ and\ \bibinfo {author} {\bibfnamefont {H.}~\bibnamefont
  {Markram}},\ }\href@noop {} {\bibfield  {journal} {\bibinfo  {journal}
  {Neural Comput.}\ }\textbf {\bibinfo {volume} {14}},\ \bibinfo {pages} {2531}
  (\bibinfo {year} {2002})}\BibitemShut {NoStop}%
\bibitem [{\citenamefont {Yildiz}\ \emph {et~al.}(2012)\citenamefont {Yildiz},
  \citenamefont {Jaeger},\ and\ \citenamefont {Kiebel}}]{Yildiz2012-mq}%
  \BibitemOpen
  \bibfield  {author} {\bibinfo {author} {\bibfnamefont {I.~B.}\ \bibnamefont
  {Yildiz}}, \bibinfo {author} {\bibfnamefont {H.}~\bibnamefont {Jaeger}},\
  and\ \bibinfo {author} {\bibfnamefont {S.~J.}\ \bibnamefont {Kiebel}},\
  }\href@noop {} {\bibfield  {journal} {\bibinfo  {journal} {Neural Netw.}\
  }\textbf {\bibinfo {volume} {35}},\ \bibinfo {pages} {1} (\bibinfo {year}
  {2012})}\BibitemShut {NoStop}%
\end{thebibliography}

\end{document}